# Counter-propagating Entangled Photon Pairs from a Monolayer

Zhuoyuan Lu [1,3#], Jiri Janousek[1,3#], Syed M. Assad[2,3,4], Shuyao Qiu[1], Mayank Joshi[2,3,4], Yecheng Hu[5], Alex Y Song[5], Chuanyu Wang[1], Manuka Suriyage[1], Jie Zhao[2,3], Ping Koy Lam[2,3,4] and Yuerui Lu[1,3*]

[1]School of Engineering, College of Engineering, Computing and Cybernetics, the Australian National University, Canberra, ACT, 2601, Australia

[2]Research School of Physics, College of Science, the Australian National University, Canberra, ACT, 2601, Australia

[3]ARC Centre for Quantum Computation and Communication Technology, the Australian National University, Canberra, ACT, 2601, Australia

[4]Quantum Innovation Centre (Q.InC), Agency for Science Technology and Research (A*STAR), 2 Fusionopolis Way, Innovis #08-03, Singapore 138634, Republic of Singapore

[5]School of Electrical and Computer Engineering, the University of Sydney, Sydney, New South Wales 2006, Australia

\# These authors contributed equally to this work.
\* To whom correspondence should be addressed: Yuerui Lu (yuerui.lu@anu.edu.au)

**Abstract:**

**Non-phase-matched spontaneous parametric down-conversion (SPDC) in atomically thin materials provides new degrees of freedom and enhanced quantum information capacity compared to conventional phase-matched sources[1]. These systems emerged as promising platforms for quantum computing[2,3], communication[4,5], and imaging[6], with the potential to support higher-order nonlinear processes. However, direct observation of photon-pair emission from a monolayer has remained experimentally challenging[7,8]. In this work, we theoretically modeled SPDC emission across the full angular space from a monolayer GaSe film and experimentally validated the model through measurements of both co- and counter-propagating photon pairs. We demonstrated two-photon quantum correlations**

**in the telecom C-band from the thinnest SPDC source reported to date. The spatially symmetric, broadband emission predicted by theory was confirmed experimentally. Furthermore, we observed high-fidelity Bell states in the counter-propagating configuration, marking the first realization of polarization-entangled photon pairs from a monolayer. Our results revealed the emission characteristics of SPDC in the deeply subwavelength, non-phase-matched regime, and introduced atomically thin, counter-propagating SPDC as a scalable and integrable platform for programmable quantum state generation, extendable via moiré superlattice engineering.**

**Keywords:** SPDC, monolayer, GaSe, frequency-angular spectrum, counter-propagating SPDC, entangled photon pairs, Bell states.

## Introduction

Entangled photon source generated from spontaneous parametric down conversion (SPDC) is one of the most common source used in quantum communication[9,10]. Improving the quantum information capacity of SPDC sources is crucial for advancing scalable quantum communication networks[5,11,12]. The quantum capacity of SPDC sources is critically influenced by factors of bandwidth and spatial emission ranges. However, conventional SPDC processes are limited by strict phase-matching requirements, which restrict both the emission bandwidth and the range of emission angles[1]. Recently, there has been growing interest in non-phase-matched SPDC processes, as these allow for much larger phase mismatches, resulting in broader bandwidths and potentially greater quantum information capacity[1,13]. Theoretical models indicate that by using ultra-thin films for SPDC sources, emission angles could surpass traditional limits, enabling photon pairs to be emitted over a full 4π solid angle, vastly expanding the emission range[7,14]. This 4π emission characteristic enables SPDC to simultaneously support co-propagating, back-propagating, and counter-propagating photon pairs, providing greater flexibility in emission control.

While counter- and back- propagating SPDC has been demonstrated in periodically poled nonlinear crystals and waveguides[15-18], as well as engineered metasurfaces[19-21], these approaches rely on fundamentally different physical mechanisms. In periodically poled structures, quasi-phase matching (QPM) is employed to compensate for phase mismatch in a nonlinear medium, enabling directional biphoton emission dictated by the material's dispersion and poling period. In contrast, metasurfaces harness subwavelength nanostructures to locally enhance nonlinear interactions and precisely control emission angles[22].

Two-dimensional (2D) materials have gained attention as promising platforms for SPDC. In particular, monolayer systems offer atomic-scale thickness and flexible interlayer control, making them attractive for integration with micro cavities, photonic fibers or to form van der walls heterostructures[23,24]. Early studies investigated SPDC in monolayer transition metal dichalcogenides (TMDs), but no unambiguous SPDC signals were detected due to significant photoluminescence (PL) background and low photon count rates[7,8]. Subsequent advancements in materials science revealed that rhombohedral (3R)-stacked vdW structures could maintain broken inversion symmetry even with increasing layer numbers, enabling efficient second-harmonic generation (SHG) in multilayered forms[25]. Unlike 2H stacking, where adjacent layers undergo a 180° rotation that restores global inversion symmetry and thereby suppresses second-order nonlinear optical processes, 3R stacking exhibits a translational shift along the c-axis without such rotational symmetry[26,27]. Following these developments, SPDC was successfully demonstrated in $NbOCl_2$ thin film (down to 46 nm thick with potential resonant enhancement)[28], as well as in 3R-stacked subwavelength thickness $MoS_2$ (294 nm thick)[29] and $WS_2$ (350 nm thick)[30], marking a significant step forward for SPDC in layered 2D materials.

However, two main challenges limit SPDC observation in further thinner platforms. First, although 2D materials like 3R-TMDs exhibit second-order nonlinear susceptibilities significantly higher than traditional nonlinear crystals (e.g., BBO and lithium niobate), the absolute conversion efficiency remains much lower in ultra-thin films than in bulk crystals[31]. Second, PL emissions from 2D layers thin film can generate strong backgrounds that obscure the SPDC signal. These emissions are sometimes defect-related and vary widely in wavelength, making them unpredictable and difficult to filter[7,32,33]. Gallium selenide (GaSe) monolayers present a potential solution to these challenges. SHG measurements and earlier studies have shown that GaSe monolayers exhibit nonlinear responses 1–2 orders of magnitude stronger in the NIR region than TMD monolayers[34]. Unlike TMDs, GaSe transitions to an indirect bandgap structure in films as thin as four atomic layers, rendering PL emissions from GaSe monolayers negligible (Supplementary Fig. 1)[35]. This unique property makes GaSe monolayers particularly attractive for SPDC sources capable of operating across a wide spectral range, without interference from fluorescence.

In this study, we used an angular-frequency model to explore the unique phase-matching conditions for SPDC biphotons generated in a GaSe monolayer—a III–VI semiconducting metal monochalcogenide characterized by its outstanding second-order nonlinearity in NIR and scalable broken symmetry, similar to 3R-TMDs and $NbOCl_2$[34]. Our experiments demonstrated both co-propagating and counter-propagating SPDC processes within the C-band telecom range produced by the GaSe monolayer, highlighting spatial emissions under deeply subwavelength, non-phase-matched regime. To further characterize the quantum properties of these biphotons, we conducted spectral analysis and polarization entanglement measurements on the counter-propagating SPDC pairs through quantum state tomography (QST). Our results confirmed the generation of maximally entangled Bell states with fidelities of 0.82 and 0.76, respectively. These findings highlight the promise of GaSe monolayers as compact, integrated sources for quantum photonics, effectively overcoming traditional phase-matching limitations and enabling robust, tunable biphoton generation in ultra-thin platforms for advanced quantum information technologies.

## Results and Discussion

**Theoretical analysis of spatially symmetric SPDC from a monolayer**

In traditional thick SPDC sources, precise control over sample thickness and wavelength is essential for minimizing phase mismatch and achieving high efficiency. In contrast, sub-wavelength SPDC sources take advantage of their limited interaction lengths, which restrict the waves from propagating sufficiently to accumulate significant phase differences and destroy the signals[36]. Consequently, as the interaction length decreases in nonlinear materials, the allowable longitudinal mismatch increases, as depicted in Fig. 1a. The enhanced tolerance for phase mismatch not only facilitates the generation of signal and idler photons over a broader spectral range—leading to increased bandwidth—but also allows photon pairs to be emitted across a wider range of angles, producing a broader angular spectrum rather than being confined to a narrow emission cone[1,37].

Our theoretical predictions use a well-established model from previous research, including the foundational description by Okoth et al.[38], as well as the framework detailed by Davoyan et al[39] (Supplementary Note 1). In a simplified form, the probability of photon-pair emission in a medium of length $L$ with a nonlinear susceptibility $\chi^{(2)}$ is given by

$$R \propto \chi(2)^2 L^2 F_{pm}(\Delta k_\parallel) F_p(\Delta k_\perp) \tag{1}$$

where $F_{pm}(\Delta k_\parallel)$ is the phase-matching function and $F_p(\Delta k_\perp)$ is the pump function. This model assumes a Gaussian pump beam with a waist radius $w_0$ propagating in the longitudinal direction. The signal and idler wavevectors can be represented as $k = k_\parallel + k_\perp$, where $k_\parallel$ is the component parallel to the pump direction, and $k_\perp$ is the component perpendicular to the pump propagation. The phase-matching function, $F_{pm}(\Delta k_\parallel) = \text{sinc}^2(\Delta k_\parallel L/2)$, controls the spectral width and angular width of the possible photon-pair emission. In contrast, the pump function, $F_p(\Delta k_\perp) = \exp[-(\Delta k_\perp w_0)^2/2]$, depends on the transverse wavevector mismatch $\Delta k_\perp = (k_{s\perp} + k_{i\perp})$ and defines the correlations between the signal and idler angles of emission as the energy conservation is preserved.

For sample with ultra-thin interaction lengths $L$, the longitudinal wavevector mismatch $\Delta k_\parallel = (k_p - k_{s\parallel} - k_{i\parallel})$, where $k_p$ is the pump wavevector, is allowed to become large, effectively relaxing phase-matching requirements and approaching a non-phase-matched regime. This leads to an exceptionally broad frequency spectrum, in stark contrast to the significantly narrower spectrum characteristic of macroscopically thick, phase-matched SPDC crystals[1]. In this extreme regime, photon-pair emission is entirely governed by the pump function $F_p(\Delta k_\perp)$. The calculated angular-frequency spectrum for monolayer GaSe (Fig. 1b) reveals key insights into photon-pair correlations, demonstrating that the SPDC emission is ultra broadband and spatially symmetric, with equal likelihood in both forward and backward directions.

Further exploration of the angular emission characteristics of signal and idler photons provides deeper understanding. Fig. 1c and Fig. 1d present the calculated angular emission spectra for signal (pink) and idler (blue) photons, integrated across the entire frequency bandwidth, for a monolayer GaSe sample and a 216-layer GaSe sample, respectively. The cross-sections of the 3D emission spatial profiles for the two samples (Supplementary Fig. 2) illustrate the number of signal and idler photons emitted within sectors defined by infinitesimally small arc angles. These plots reveal that the probabilities of signal and idler emission vary with angle, a behavior governed by material dispersion and influenced by the interplay between phase-matching and pump functions. It is evident that as the thickness of the sample increases, the dominance of emission in the forward direction becomes more pronounced.

To clarify the underlying physical process, we plot the signal and idler emission of monolayer as a function of the emission angle, integrated over an experimental frequency bandwidth (Fig. 1e). Four possible photon-pair generation scenarios emerge: (I) both the signal and idler photons emitted in the forward direction, (II) the signal emitted in the forward direction with the idler in the backward direction, (III) the reverse of scenario II, and (IV) both photons emitted in the backward direction. For a monolayer, each scenario occurs with equal probability. Notably, when performing a correlation measurement by detecting counter-propagating photons, the likelihood of detecting photon pairs is twice as high as when measuring in either the forward or backward direction alone. This factor of 2 arises from the simultaneous detection of forward-backward propagating single-idler and idler-signal photon pairs. As sample thickness increases, however, this symmetry gradually breaks. For instance, in the 216-layer sample shown in Fig. 1f (with angular-frequency spectrum calculated in Supplementary Fig. 3), photon pairs are more likely generated in the forward (co-propagating) direction, reducing the counter-to-co-propagating pair ratio to less than 0.25. This result shows that emission symmetry can break down even well below the coherence length. At 1550 nm, GaSe's coherence length is around 3.5 µm, yet a 173 nm (216-layer) sample already shows significant asymmetry. At shorter excitation wavelengths such as 405 nm, stronger dispersion reduces the coherence length to <200 nm, making the spatial emission more thickness sensitive (Supplementary Fig. 4b–d). Simulations indicate that only monolayer or few-layer GaSe can retain near-ideal spatial symmetry in the visible regime. Therefore, under short-wavelength excitation and in applications demanding high spatial indistinguishability of photon pairs—such as quantum imaging[40,41], high-dimensional quantum communication, quantum key distribution[42], and quantum metrology[41,43,44]—monolayer GaSe becomes a particularly attractive platform.

**Experimental demonstration of bidirectional SPDC**

To experimentally validate these concepts, we first prepared a sample of GaSe monolayer on fused silica, an optically transparent substrate spanning the visible to near-infrared range, as shown in Fig. 2a (details in Methods). Then we built an SPDC setup based on the Hanbury-Brown and Twiss topology (Fig. 2b; details in Methods and Supplementary Note 2) capable of measuring both co-propagating and counter-propagating photons from GaSe monolayer. The setup is carefully optimized to ensure equal optical loss in both scenarios, enabling a direct and unbiased comparison of photon-pair emission from the GaSe monolayer. With only two SPADs available, measurements for each scenario were taken separately: both detectors were placed in the forward direction for the co-propagating measurement, while one detector was placed forward and the other backward for the counter-propagating photon-pair detection. As a potential extension of this system, we further proposed and simulated a fiber-integrated platform, where a monolayer GaSe is positioned at the shared focal plane of two opposing Fresnel lenses. This design not only preserves the intrinsic forward-backward symmetry of SPDC emission but also enables direct coupling of photon pairs into optical fibers, offering enhanced collection efficiency and integrability. The full simulation results and structural schematics are presented in Supplementary Note 3 and Supplementary Fig. 5.

We measured correlations for both scenarios as a function of pump power (Fig. 2c). As predicted by theory, this dependence is linear with pump power and the slopes of the fitted curves reveal a twofold increase in the probability of photon-pair emission in the counter-propagating configuration compared to the co-propagating one. This resulted in maximum coincidence rates, corrected for the optical losses from the setup and detector efficiency (Supplementary Note 4), of 8.4 Hz for the counter-propagating scenario and 4 Hz for the co-propagating scenario at a pump power of 40 mW.

To verify generation of non-classical correlated photon pairs, we performed the $g^{(2)}(\tau)$ second-order correlation measurement. The measured $g^{(2)}(\tau)$ functions for a monolayer at a pump power of 40 mW (Fig. 2d) show $g^{(2)}(0)$ peaks of 1.75 for the co-propagating case and 3.3 for the counter-propagating case. Since $g^{(2)}(0)$ is positively correlated with the signal-to-noise ratio under low photon flux conditions, the reduced photon-pair generation probability and the higher background noise for the both forward-direction-positioned detectors (e.g. residual pump scattering and Raman noise) in the co-propagating case naturally leads to a smaller $g^{(2)}(0)$ value. While the $g^{(2)}(0)$ value of 1.75 for the co-propagating scenario is slightly below the classical threshold of 2, it still indicates the presence of photon-pair correlations, This reduction is primarily due to the relatively low coincidence rate from monolayer GaSe compared to thicker samples. In contrast, the $g^{(2)}(0)$ value of 3.3 in the counter-propagating case provides strong evidence for SPDC generation from a GaSe monolayer in this experimental setup. Notably, this result marks, to our knowledge, the first experimental observation of photon-pair generation from a monolayer vdW material—an achievement that addresses a long-standing challenge in the field[7,8]. For comparison, we also measured the counter-to-co-propagation coincidence ratio for a 216-layer GaSe sample (Fig. 2e), confirming our theoretical predictions that photon pairs are predominantly generated in the forward direction. The measured $g^{(2)}(\tau)$ functions for both scenarios are provided in Supplementary Fig. 6.

To showcase the remarkably broad photon-pair emission bandwidth, we performed an SPDC spectrum measurement. To maximize detection efficiency of the setup, we removed the 50/50 beam splitter and configured the system in a counter-propagating arrangement. To control the wavelengths detected, we placed fiber-coupled tunable filters (details in Methods) before each detector, setting each measurement at a specific wavelength with a bandwidth of 10 nm. Due to detector bandwidth limitations, we were only able to record coincidences in the 1460–

1650 nm range. However, as shown in Fig. 2f, this represents just a small portion of the total emission bandwidth predicted by our model. Theoretical calculations—based on our experimental collection angle—indicate a frequency bandwidth of approximately 100 THz (Supplementary Fig. 4a). Assuming a Lorentzian emission profile, this corresponds to a biphoton temporal correlation time of 3.2 fs, opening opportunities for advanced applications in ultrafast quantum optics and quantum information science[45-47].

**Counter-propagating entanglement from GaSe monolayer**

The GaSe retains its $D^1_{3h}$ symmetry group from monolayer to bulk crystal form[34]. In this symmetry, four non-zero tensor elements contribute to the second-order susceptibility, with the relationship $\chi^{(2)}_{\alpha\beta\gamma} = \chi^{(2)}_{yyy} = -\chi^{(2)}_{yxx} = -\chi^{(2)}_{xxy} = -\chi^{(2)}_{xyx}$, where $x$ and $y$ are the in-plane Cartesian coordinates representing the polarizations of the SH and fundamental fields. As Klimmer et al. noted[48], these $x$ and $y$ directions correspond to the zigzag (ZZ) and armchair (AC) orientations of the crystal lattice, respectively. A six-fold symmetric, polarization-dependent SHG response was observed in GaSe (see Supplementary Fig. 7). In our prior work, we demonstrated co-propagating Bell-state generation by exploiting these non-vanishing nonlinear susceptibilities in 3R-MoS$_2$[29]. As shown in Fig. 3a, by adjusting the pump polarization along the AC and ZZ orientations of the crystal lattice, we can generate Bell states of different forms: with pump polarization along the AC orientation, we produce the state $|\Phi^-\rangle = 1/2(|HH\rangle - |VV\rangle)$, and along the ZZ orientation, we produce $|\Psi^+\rangle = 1/2(|HV\rangle + |VH\rangle)$.

Building on this knowledge, we extended our approach to monolayer GaSe, achieving Bell-state generation for the first time in a monolayer vdW material. To enable this, we modified the SPDC setup to operate in the counter-propagating configuration and incorporated two sets of polarization optics for QST (Fig. 3b, details in Methods). The density matrix $\hat{\rho}$ for the quantum states in both the AC and ZZ directions was reconstructed using maximum likelihood estimation, based on 16 required measurements as detailed in Supplementary Note 5[3,49]. Given the low photon-pair production rate, we recorded each measurement over a 2-hour period to accumulate sufficient statistics. The resulting fidelity matched closely with ideal Bell states, yielding fidelities $F_{AC} = 0.82$ and $F_{ZZ} = 0.76$ (concurrence $C_{AC} = 0.82$ and $C_{ZZ} = 0.59$) for the $|\Phi^-\rangle$ (Fig. 3c) and $|\Psi^+\rangle$ (Fig. 3d) Bell states, respectively. These slightly reduced fidelities are mainly due to detector dark counts level. We further verified that higher fidelities can be achieved in this experimental configuration by using thicker samples, which yield a higher ratio of true photon-pair coincidences to accidental coincidences. For instance, using a 90-layer thick GaSe sample (Supplementary Fig. 8), we achieved fidelities $F_{AC} = 0.98$ and $F_{ZZ} = 0.98$ (concurrence $C_{AC} = 0.97$ and $C_{ZZ} = 0.98$), affirming the precision of our setup and the performance of the optical components used.

**Conclusion**

In conclusion, we extended the established microscale theory of non-phase-matched SPDC to the nanoscale and applied it to model the spatial emission dynamics in monolayer films. Experimentally, we validated the spatially symmetric photon-pair emission under the non-phase-matched regime in a monolayer GaSe within the telecom C-band. Additionally, we demonstrated tunable polarization-entangled Bell states in the counter-propagating geometry, marking—to the best of our knowledge—the thinnest entangled photon-pair source reported to date.

The broadband $4\pi$ emission, alongside high entanglement fidelity, establishes this platform as a promising source for high-dimensional photonic states. Future advances in materials—such as rBN, SnP$_2$Se$_6$[50-52], and moiré-engineered heterostructures—may further extend the operational range and enable complex entanglement states and photon indistinguishability[53-56].

While the current efficiency is limited by the monolayer's sub-nanometer interaction length, several photonic strategies have emerged to enhance brightness, including cavity coupling[57-59], lens integration, waveguide integration[60,61], and twist-phase matching via twisted stacking[62-64]. These developments collectively suggest a viable route toward bright, scalable, and integrable quantum light sources based on monolayer.

**Materials and Methods**

*Material fabrication and thickness characterization*: GaSe crystals (from 2D semiconductors) were synthesized using the Bridgman method and subsequently mechanically exfoliated onto 500 μm fused silica substrates, which exhibit low dielectric screening. Supplementary Fig. 9 details the dielectric screening effects of various substrate and encapsulation schemes, which are consistent with our previous studies[65]. Characterization of the samples was conducted using PL spectroscopy with a Horiba LabRAM system. For GaSe thin films, a Bruker MultiMode III AFM was employed to determine their thickness.

*Spontaneous parametric down-conversion:* An amplified 1550 nm single-mode CW laser was directed into a resonant bow-tie cavity containing a 15 mm long periodically poled potassium titanyl phosphate (PPKTP) crystal. This setup generated 775 nm light through SHG, which was then fiber-coupled and used as the pump source for SPDC experiments.

In the setup for analyzing co- and counter-propagating SPDC in monolayer GaSe (additional details in Supplementary Note 2), the 775 nm pump beam was focused onto the GaSe sample through a high-NA aspheric lens, achieving a 10 μm beam waist. Forward propagating SPDC photons were collected with an identical aspheric lens and the pump laser was blocked by two long-pass filters. A broadband 50/50 beam splitter was used for correlation measurements, with both reflected and transmitted beams coupled into a pair of multimode fibers. Backward propagating SPDC photons were collected through the same aspheric lens used for focusing the pump, reflected off the dichroic mirror, filtered from the residual pump, and coupled into a multimode fiber using identical components. For photon detection, two SPADs were used, and the time-correlated single-photon counting (TCSPC) system was connected to both SPADs to process the photon pair correlations. Note that the goal of our study was not to investigate the spatial correlations of SPDC photon pairs. Instead, we measured the emission rate of photon pairs emerging from the sample within collection cones defined by the detection angle $\theta_{det} = NA/n_{GaSe}$, using lenses with identical numerical apertures to collect both co- and counter-propagating photons.

For frequency spectrum and QST measurements, we employed a counter-propagating setup, doubling photon-pair detection efficiency compared to the co-propagating regime (see main text). This setup excluded the 50/50 beam splitter, coupling all forward-propagating photons into a single multimode fiber. This configuration maximizes efficiency and accommodates the use of only two SPAD detectors available for the experiment. Tunable fiber-coupled bandpass filters, set at 10 nm bandwidth, were used for spectral measurements. Polarization entanglement was assessed via QST using quarter-wave plates, half-wave plates, and broadband linear polarizers in each arm.

*Polarization entangled quantum state tomography:* For QST, we pumped along the crystal axis of the sample, either in the armchair or zigzag direction. Using polarization-dependent SHG measurements (details in Supplementary Note 6), we first identified the crystal axis, then adjusted the pump polarization accordingly. For the SPDC photons, the half-wave and quarter-wave plates were set to align the crystal axis polarization with the output polarizer. We performed QST by measuring each photon of the SPDC pair in the H, V, D, A, R, and L polarization bases, selecting 16 combinations to reconstruct the entangled state's density matrix $\rho$. The reconstruction was carried out using maximum likelihood estimation, as detailed in Supplementary Note 5 and following the methodology established in our previous work[29].

**Acknowledgements**

The authors acknowledge funding support from ANU PhD student scholarship, Australian Research Council (grant no. DP240101011, DP220102219, LE200100032) and ARC Centre of Excellence in Quantum Computation and Communication Technology (project number CE170100012) and the National Health and Medical Research Council (NHMRC; ID: GA275784).

**Competing interests**

The authors declare that they have no competing financial interests or any other conflict of interest.

**Author Contributions**

Y. L. conceived and supervised the project; Z.L prepared the samples; J.J and Z.L carried out the optical measurements; J.J. and S.A. conduct the simulation; Z.L., J.J and Y.L. analyzed the data; Z.L., J.J. and Y. L. drafted the manuscript, and all authors contributed to the manuscript.

**Supplementary Information**

All additional data and supporting information and methods are presented in the supporting information file available online. The figures and information in the supporting information have been cited at appropriate places in this manuscript.

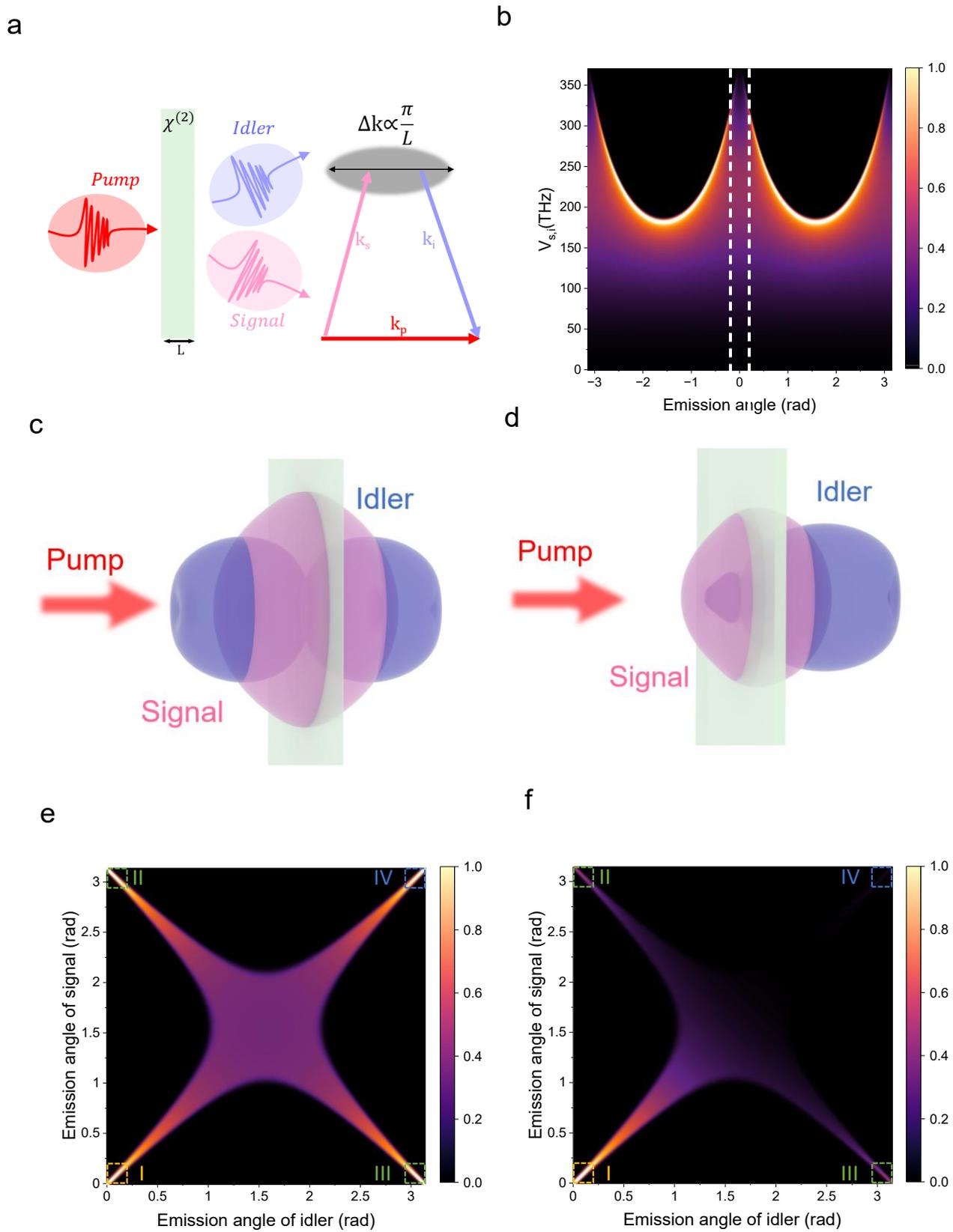

Figure 1

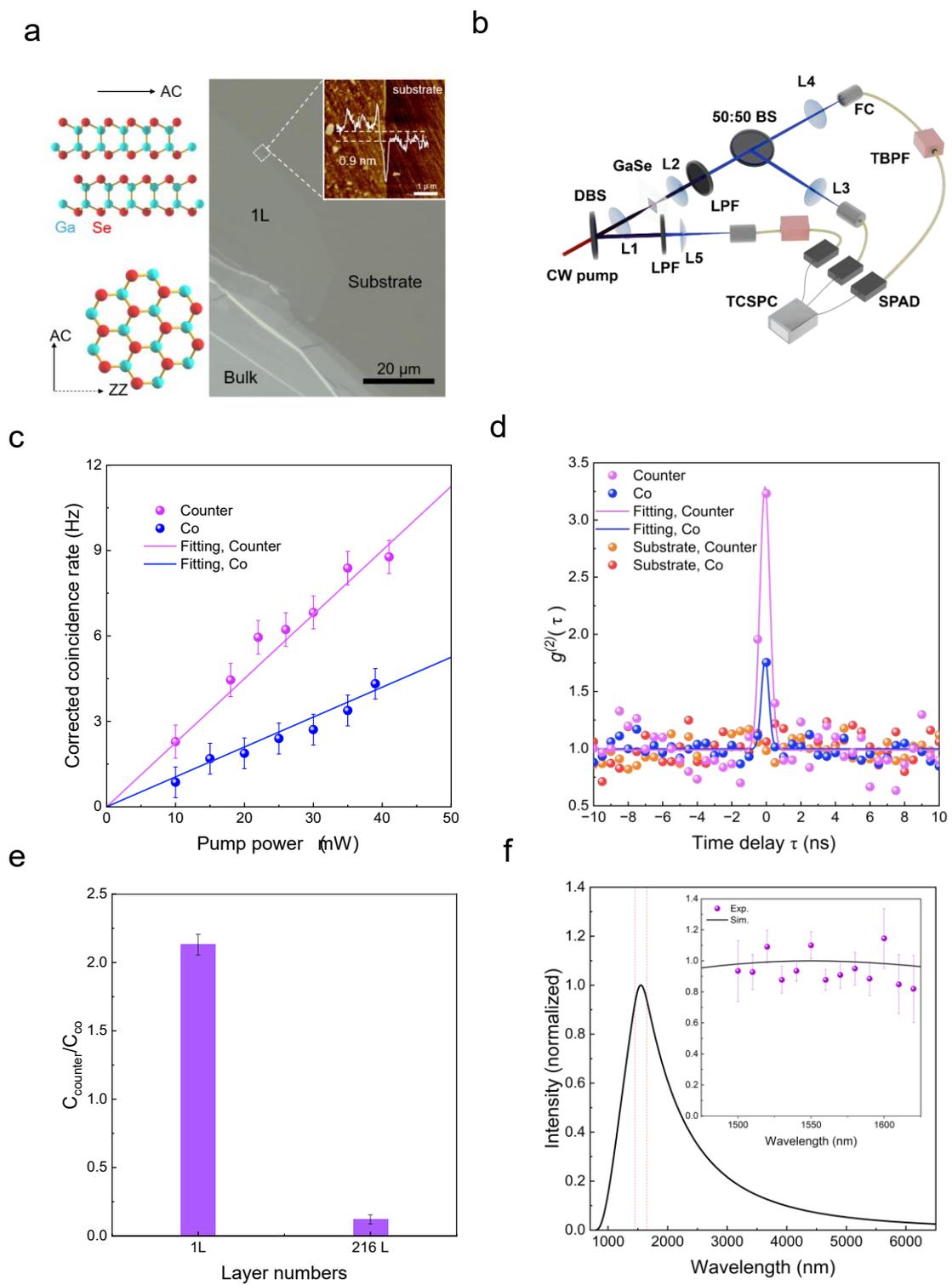

Figure 2

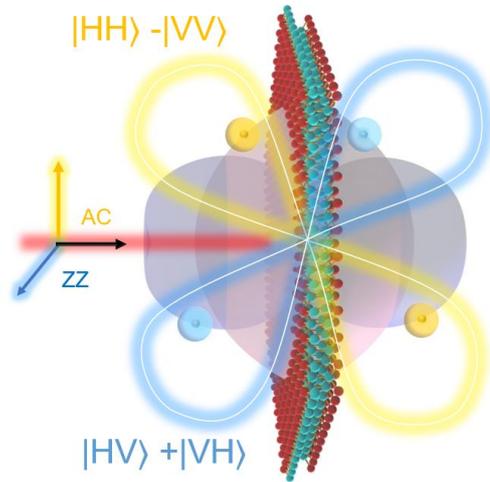
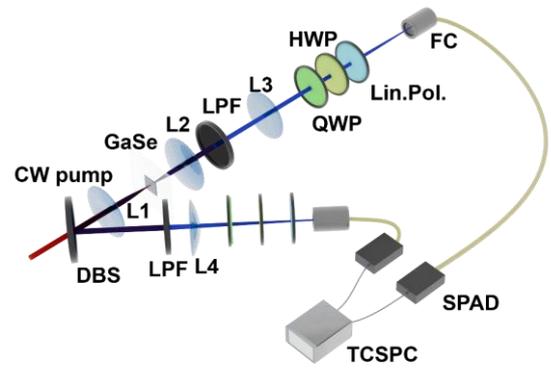
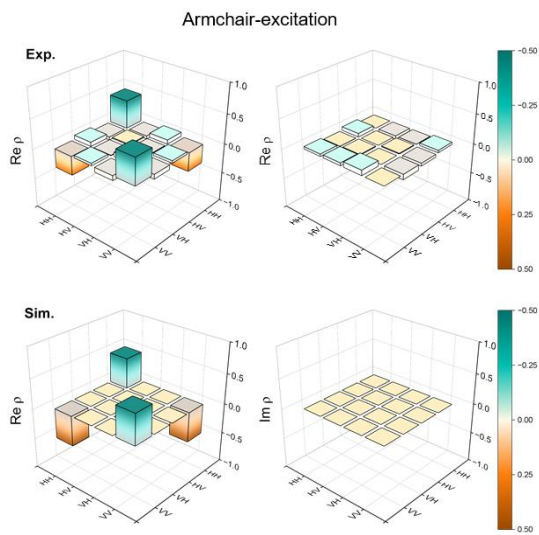
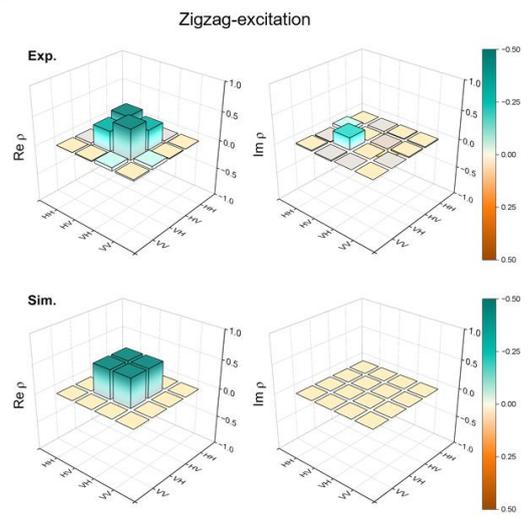

Figure 3

# FIGURE CAPTIONS

**Figure 1 │ Theoretically predicted symmetric probability of photon-pair generation from a monolayer via the non-phase matched SPDC. a**, Conceptual illustration showing the SPDC process in a $\chi^{(2)}$ nonlinear medium with thickness $L$ (left) and the allowed wavevector mismatch $\Delta k$ inversely proportional to $\pi/L$ (right). **b**, Calculated frequency-angular spectrum of non-phase-matched SPDC from a GaSe monolayer, with dashed lines depicting the experimental collection angle in the forward direction. **c-d**, Simulated spatial angular emission profiles of the signal and idler photons from a monolayer (c) and a 216-layer GaSe sample (d), respectively, with the red arrow indicating the direction of the pump laser. The light green sheets represent the samples. For the monolayer, the emission profile is symmetric, producing SPDC photons equally in the forward and backward directions. However, as the thickness of the nonlinear medium increases, the SPDC emission becomes predominantly forward-directed, reflecting the influence of increased material thickness on phase-matching conditions. **e-f**, Simulated signal and idler emissions from a monolayer (e) and a 216-layer GaSe sample (f), respectively, plotted as a function of emission angles and integrated over the experimental frequency bandwidth. The dashed boxes, each with a size of 0.2 radians, represent the collection angles for the four SPDC scenarios detailed in the main text.

**Figure 2 │ Experimental demonstration of co- and counter-propagating SPDC from a GaSe monolayer. a**, Left: Van der Waals molecular structure of GaSe depicted from both top and side views, with the coordinate system defined by the armchair (AC) and zigzag (ZZ) directions of the crystal. Right: Optical microscope image of a monolayer GaSe sample exfoliated on fused silica substrate. The inset, a magnified view of the dashed box, shows the AFM image and corresponding thickness profile of the monolayer region. **b,** Experimental setup for co- and counter-propagating SPDC and the spectrum measurement (details in Methods and Supplementary Note 2). CW laser pump at 775 nm. The monolayer and bulk GaSe flakes were exfoliated onto fused silica substrates. DBS: dichroic beam-splitter, BS: broadband 50/50 beam-splitter, L: mode-matching lens, LPF: long-pass filter, TBPF: tunable band-pass filter, FC: fibre coupler, SPAD: single-photon avalanche diode detector, TCSPC: time-correlated single-photon counting module. **c**, Measured pump-power-dependent

coincidence rates from a monolayer GaSe sample in both co- and counter-propagating configurations, within a detection bandwidth of 1460–1650 nm and an experimental collection angle of 0.2 radians. The rates have been corrected for optical losses in the setup and for detector efficiency. **d**, Measured second order correlation function $g^{(2)}(t)$ from a monolayer GaSe sample in co-propagating and counter-propagating scenarios, with average excitation power $P$ = 40 mW and integration time 3 hours. **e**, Measured coincidence ratio between counter- to co-propagating SPDC from a monolayer and a 216-layer GaSe sample. **f,** Simulated and measured (inset) SPDC bandwidth from a GaSe monolayer, with the SPDC emission rate at 1550 nm normalized to 1. The red dashed lines indicate the experimentally measured wavelength range (1460–1650 nm), as shown in the inset. The calculated bandwidth spans 900 nm, corresponding to approximately 100 THz.

**Figure 3 | Counter-propagating polarization-entangled photon pairs from a GaSe monolayer. a,** Schematic of entangled photon pairs generated in GaSe monolayer, with the two Bell states $|\Phi^-\rangle = 1/2(|HH\rangle - |VV\rangle)$ and $|\Psi^+\rangle = 1/2(|HV\rangle + |VH\rangle)$ corresponding to the pump polarization aligned with armchair (AC) and zigzag (ZZ) directions, respectively. **b,** Experimental setup for counter-propagating QST. HWP: half wave plate, QWP: quarter wave plate, Lin. Pol.: linear polarizer. **c-d**, Experimentally measured (top) and simulated (bottom) Bell states for AC (c) and ZZ (d) pump excitations, respectively. The quantum state density matrix, $\hat{\rho}$, was reconstructed using maximum likelihood estimation, leveraging 16 measurements required for QST[3,49].